\documentclass[showpacs,twocolumn,preprintnumbers,prl,amssymb]{revtex4}
\usepackage{graphicx}
\usepackage{dcolumn}
\usepackage{bm}
\usepackage{psfrag,epsfig}

\newcommand{\Op}{{\mathcal{O}}(p)}
\newcommand{\Opd}{{\mathcal{O}}(p^2)}
\newcommand{\Opt}{{\mathcal{O}}(p^3)}

\newcommand{\be}{\begin{equation}}
\newcommand{\ee}{\end{equation}}
\newcommand{\ba}{\begin{eqnarray}}
\newcommand{\ea}{\end{eqnarray}}
\newcommand{\nn}{\nonumber}

\begin{document}

\preprint{CAFPE-63/05, UG-FT-193/05}

\title{
Surprises in threshold antikaon-nucleon physics}

\author{Jos\'e A. Oller$^1$, Joaquim Prades$^2$ and Michela Verbeni$^2$}
\affiliation{$^1$ Departamento de F\'{\i}sica, Universidad de Murcia, 
E-30071 Murcia, Spain\\$^2$ Centro Andaluz de F\'{\i}sica
de las Part\'{\i}culas Elementales (CAFPE) and 
Departamento de F\'{\i}sica Te\'orica y del Cosmos,\\ 
Universidad de Granada, 
Campus de Fuenta Nueva, E-18002 Granada, Spain}
\date{\today}

\begin{abstract}
\noindent
Low energy $\bar{K}N$ interactions are studied within 
Unitary Chiral  Perturbation Theory at
 next-to-leading order with ten coupled channels. We pay special
attention to the recent precise determination of the strong shift 
and width of the kaonic hydrogen  $1s$ state by the DEAR Collaboration 
that has challenged   our theoretical
understanding of this sector of  strong interactions. 
We typically find two classes  of solutions, both of them 
reproducing previous data,  that either can or cannot
accommodate the DEAR  measurements. The former class 
 has not been previously discussed.
\end{abstract}


\maketitle

\noindent
{\bf 1. }
Low energy antikaon-nucleon interactions have been object of 
extensive study  almost for the last 50 years. 
Based on  early data on $K^-p$ scattering, Dalitz and Tuan  predicted \cite{df1959}
in 1959  the existence of a sub-threshold $\bar{K}N$ resonance, 
the $\Lambda(1405)$, first seen experimentally three years 
later \cite{lamb}. Despite this
success, $K^-p$ scattering is still challenging our understanding 
of strong interactions.
 First, this resonance,  being too light, appears puzzling 
for quark model \cite{hamaie}  and lattice  QCD \cite{adelaida} communities.
 This fact can be interpreted as one more evidence  that the 
$\Lambda(1405)$  is a 
 dynamically generated resonance as claimed in refs.\cite{df1959,speth,uni,ollerm}. 
Second, there has been disagreement between the sign of the $K^-p$ 
scattering lengths   extracted either from scattering or 
from the $1s$  $K^-p$  atomic level shift until 1998 
 when it was settled down
 by the KpX experiment at KEK \cite{ito}.  Now, the  around  factor 
of two more precise  DEAR measurement  \cite{DEAR}
brings in a further disagreement  with all previous theoretical 
results from SU(3) chiral dynamics,
  which are however  compatible with the KEK 
measurement \cite{ollerm,martin,bura}.   Third, 
 the physical $\Lambda(1405)$ has not yet been considered 
 up to very recently \cite{ollerm,team} as the admixture of
two nearby poles, so that different reactions weighting more one pole
or the other produce different resonant shapes  
peaking at different energies.
For experimental  evidences on this issue see \cite{osetprl}. 
Fourth, the recently discovered strange tri-baryons 
$S^0(3115)$ and $S^1(3140)$ \cite{suzu} have 
 most likely shown that deeply bound states of $\bar{K}$, 
as predicted in \cite{yama} and even deeper, do exist. 
The $\bar{K}$-nucleus potential is therefore definitely strongly
  attractive in contrast with the up to now prevailing 
believes and claims of  a shallow potential. 
This is of foremost importance as it is  a way to obtain very dense 
nuclear matter \cite{yama}, $(3\sim4)\times\rho_0$, 
  as well as to get kaon condensation in nuclear matter
 (e.g. neutron stars)  \cite{manke}, or  strangeness clusters 
in heavy-ion collisions.   A sounder theoretical explanation  of such 
strongly attractive $\bar{K}$-nucleus potential is called for. 
 Fifth, there is an exhaustive
 search of the strangeness content of the proton with positive 
results in several experiments  worldwide \cite{gzero}. Furthermore, 
the recent evaluation \cite{pavan} of the 
pion-nucleon sigma term $\sigma_{\pi N}$, points toward a rather large  
 strangeness content of the proton, with a contribution to the
nucleon mass between 110 to 220 MeV. 
 One can address  this issue by calculating the  
proton scalar form factor,  $\langle p| \bar{s}s|p\rangle$, 
which by unitarity is related  with the $I=0$ S-wave  $\bar{K}N$  
amplitudes \cite{gassersigma}, the subject of this letter. 
All these issues concern our basic knowledge of strong  interactions
and  require as a necessary first step a precise understanding and 
calculation of the $\bar{K}N$ strong amplitudes, specially  at low and 
sub-threshold energies.  

In the limit of massless $u$, $d$ and $s$ quarks,  the QCD Lagrangian is 
symmetric  under the chiral group SU(3)$_L\times$ SU(3)$_R$. 
 Once this  symmetry is spontaneously broken to the diagonal 
${L+R}$ subgroup there appear  eight Goldstone bosons which acquire  mass  
proportionally  to the  non-vanishing quark masses
--pions, kaons and etas.
 Their low energies interactions are therefore fixed 
and can be cast in a Taylor expansion in powers 
of momenta and quark  masses modulated by unknown coefficients.  
This is known as Chiral Perturbation Theory (CHPT)  \cite{wein}. 
However, in  a system like $\bar{K}N$, 
where the  $\Lambda(1405)$ resonance is so  close to threshold, 
CHPT needs  to be supplied with a non-perturbative resummation scheme.
We follow here  the Unitary CHPT (UCHPT) \cite{ollerm} 
pioneered in \cite{nd}. This set up
was used in \cite{bura} to study $\bar{K}N$  scattering
as well. There, the authors were not able to
reproduce simultaneously  previous $\bar{K}N$ scattering data 
and the new precise DEAR measurement, and called for a possible 
 inconsistency  between the latter and former data. 
 We will show below that this is not the case. 
 
{\bf 2. } Meson-baryon interactions are described 
to lowest order in the CHPT expansion, i.e. at $\Op$, by
the Lagrangian
\ba
{\cal L}_1&=&\langle i\bar{B}\gamma^\mu [D_\mu,B]\rangle-m_0 
\langle\bar{B}B \rangle \nn\\
&+&\frac{D}{2}\langle \bar{B}\gamma^\mu
\gamma_5\{u_\mu,B\}\rangle +\frac{F}{2} \langle \bar{B}\gamma^\mu 
\gamma_5 [u_\mu,B]\rangle~,
\label{lag1}
\ea
where $m_0$ stands for the octet baryon mass in the chiral limit. 
The trace $\langle \cdots\rangle$ runs over flavor indices and 
axial-vector couplings are constrained by $F+D=g_A=1.26$. We use
 $D=0.80$ and $F=0.46$ extracted from hyperon decays \cite{hyper}. 
Furthermore,  $u_\mu=iu^\dagger
(\partial_{\mu} U) u^\dagger$, $U(\Phi)=u(\Phi)^2=\exp(i\sqrt{2}\Phi/f)$, 
with $f$ the pion decay constant in the chiral limit, 
and the covariant derivative $D_\mu=\partial_\mu+\Gamma_\mu$ 
with $\Gamma_\mu= [u^\dagger,\partial_\mu u]/2$. 
The $3\times 3$ flavor-matrices $\Phi$ and $B$ collect the lightest
octets of pseudo-scalar mesons $(\pi,K,\eta)$ and 
baryons $(N,\Sigma,\Lambda,\Xi)$, respectively. At  
next-to-leading order (NLO) in CHPT, i.e. $\Opd$,
the meson-baryon interactions are described by the Lagrangian
\ba
{\cal L}_2&=&b_0\langle \bar{B}B\rangle \langle 
\chi_+\rangle + b_D\langle{\bar{B}\{\chi_+,B\}}\rangle + b_F \langle
\bar{B}[\chi_+,B]\rangle\nn\\
&+& b_1\langle \bar{B}[u_\mu,[u^\mu,B]]\rangle + b_2 \langle
\bar{B}\{u_\mu,\{u^\mu,B\}\} \rangle \nn\\
&+&b_3\langle \bar{B}\{u_\mu,[u^\mu,B]\}\rangle
+b_4\langle \bar{B} B\rangle \langle u_\mu u^\mu\rangle+\cdots~.
\label{lag2}
\ea
Here ellipses denote terms that do not produce
new contributions to S-wave scattering at $\Opd$.
 In addition,  $\chi_+=u^\dagger \chi u^\dagger+u\chi^\dagger u$, 
$\chi=2B_0 {\cal M}_q$, ${\cal M}_q$ 
is the diagonal quark mass matrix $(m_u,m_d,m_s)$ and 
$B_0 f^2\equiv-\langle 0| \bar{q}q|0\rangle$ the quark
 condensate in the SU(3) chiral limit. 
The $b_i$ are  fitted to data.
 
 {\bf 3.} We evaluate within CHPT at $\Opd$ all two-body  
scattering amplitudes with strangeness $S$=$-1$ corresponding to the ten
coupled channels: $\pi^0 \Lambda$,
$\pi^0\Sigma^0$, $\pi^-\Sigma^+$, $\pi^+ \Sigma^-$, $K^- p$, 
 $\bar{K}^0 n$, $\eta \Lambda$, $\eta
\Sigma^0$, $K^0 \Xi^0$ and $K^+ \Xi^-$, in increasing threshold 
energy order. Each channel is labeled by its position (1 to 10) 
in the previous list. 
We denote the CHPT amplitudes at $\Op$  by 
$T_\chi^{(1)}\,\!\!_{ij}$
and  at  $\Opd$ by $T_\chi^{(2)}\,\!\!_{ij}$, with 
subindices $ij$ indicating the scattering process $i\rightarrow j$. 
We employ these perturbative amplitudes, given explicitly 
in \cite{opv2}, as input to  UCHPT at NLO.
Two-body partial wave amplitudes 
can be written in matrix notation as  \cite{ollerm}:
\be
T(W)=\left[{I}+{\cal T}(W)\cdot g(s)\right]^{-1}\cdot{\cal T}(W)~,
\label{u1}
\ee
with $W$ the total energy in the center 
of mass (CM) frame and $s=W^2$. This equation was
  derived in \cite{ollerm} employing a coupled channel dispersion relation for
  the inverse of a partial wave $T_{ij}$.
 The unitarity or right hand cut 
is taken into account easily  by the discontinuity of $T^{-1}(W)$ 
for $W$ above the $i_{th}$ threshold, 
given by the phase space factor $\delta_{ij}q_i/8\pi W$, 
with $q_i$ the CM tri-momentum.
This  is included in the diagonal matrix $g(s)$ where $g(s)_i$ 
is the  $i_{th}$ channel unitarity bubble. 
The dispersion relation above is once subtracted 
so that we introduce a subtraction constant
$\widetilde a_i$ for each channel in the $g(s)_i$ function. 
 In our problem, isospin symmetry reduces the number of subtraction constants
from 10 to 6 \cite{team}, $\widetilde a_1$, 
$\widetilde a_2=\widetilde a_3=\widetilde a_4$, 
$\widetilde a_5=\widetilde a_6$, $\widetilde a_7$, $\widetilde a_8$ 
and $\widetilde a_9=\widetilde a_{10}$. 
Our $\widetilde a_i$ satisfy $\widetilde a_i\equiv a_i(\mu)-2 \log \mu +1$, with $a_i(\mu)$ 
the subtraction constants in \cite{ollerm}.
 The  interacting kernel ${\cal T}(W)$ in (\ref{u1})
is a  $10\times 10$ symmetric matrix incorporating 
local and pole terms as well as crossed channel dynamics contributions
in the dispersion relation for $T^{-1}$.
 The matrix ${\cal T}$ (${\cal T}={\cal T}_1+{\cal T}_2+\cdots$,
 where  subindices indicate the chiral order) 
is fixed by matching (\ref{u1}) with
 the baryon CHPT amplitudes $T_\chi$ 
order by order \cite{ollerm}. At leading order, $\Op$, 
 ${\cal T}_1=T_\chi^{(1)}$ \cite{ollerm}
while at NLO, $\Opd$, ${\cal T}_2=
T_\chi^{(2)}$. The matching can be done to any arbitrary 
order and for $\Opt$ or  higher ${\cal T}_n\neq T_{\chi}^{(n)}$. 
 
{\bf 4. }  The data we include in our fits are
the $\sigma(K^- p\rightarrow K^- p)$  elastic  cross section, 
the charge exchange one, 
 $\sigma(K^- p\rightarrow \bar{K}^0n)$,
and several hyperon production reactions, 
$\sigma(K^- p\rightarrow \pi^+\Sigma^-)$, 
$\sigma(K^- p\rightarrow \pi^-\Sigma^+)$,  
$\sigma(K^- p\rightarrow \pi^0\Sigma^0)$ 
and  $\sigma(K^- p\rightarrow \pi^0\Lambda)$. 
In addition, we also fit the precisely measured ratios 
at the $K^- p$ threshold:
\ba
\gamma&=&\frac{\sigma(K^-p\rightarrow \pi^+\Sigma^-)}
{\sigma(K^-p\rightarrow \pi^-\Sigma^+)}=2.36\pm 0.04~,\\
R_c&=&\frac{\sigma(K^-p\rightarrow \hbox{charged particles})}
{\sigma(K^-p\rightarrow \hbox{all})}=0.664\pm0.011~,\nn\\
R_n&=&\frac{\sigma(K^-p\rightarrow \pi^0\Lambda)}
{\sigma(K^-p\rightarrow \hbox{all neutral states})}=0.189\pm 0.015,
\nn
\label{ratios}
\ea
see \cite{ollerm} for references. The first two ratios,
being Coulomb corrected, are measured with 1.7$\%$ precision, 
i.e, of the same order as  the expected isospin violation 
which neither we do fully consider here  nor was in \cite{bura}. 
Indeed, all the other observables  we fit have
uncertainties larger than $5\%$.
 Since we just include S-wave  amplitudes and P-waves start to contribute 
 at higher momenta \cite{opv2}, we  only include in the
fit the $K^-p$ cross  sections  low energy data points, namely, 
with  laboratory frame   $K^-$ tri-momentum  $p_L\leq 0.2$ GeV.
 This also enhances 
the sensitivity  to the lowest energy region in which  we are particularly 
interested  and where UCHPT  is more suitable.  We also 
 include in the fits the $\pi^{\pm}\Sigma^{\mp}$ 
event distributions from \cite{hemingway}  in average
--this  largely eliminates  the $I=1$ contribution.
 To calculate them we follow \cite{ollerm}.
 The number of data points included in each fit  without 
the DEAR data is 94. 
Unless the opposite is stated, we also include  in the fits 
the DEAR \cite{DEAR} measurement of the shift and width of the 
$1s$ kaonic hydrogen level
\ba
 \Delta E&=&193\pm 37 (stat) \pm 6 (syst.) \hbox{ eV},\nn\\
\Gamma&=&249 \pm 111 (stat.)\pm 39 (syst.) \hbox{ eV}\, , 
\label{deardata}
\ea 
which is around a factor two more precise than  the KEK \cite{ito}
measurement,
$\Delta E=323 \pm 63\pm 11$ eV and $\Gamma=407\pm 208 \pm 100$ eV.
To calculate the shift and width of the $1s$  
kaonic hydrogen state we use the results in \cite{akaki}
incorporating  isospin  breaking corrections. 
We compare them with the ones from the Deser formula 
\cite{deser}.  In addition, we keep the physical 
masses of mesons and  baryons in the calculation of 
 $g(s)_i$ which produces pronounced cusp effects.
  We further constraint our fits by computing   
several $\pi N$  observables calculated in baryon SU(3) CHPT 
at $\Opd$  with the values of the low energy
 constants  determined in the fit. 
Unitarity corrections in the $\pi N$ sector
are  not as large as in the 
 $S=-1$ sector  and hence a calculation within 
pure SU(3) baryon CHPT  is more reliable. Thus,  we calculate    
  $a_{0+}^+$, the isospin-even S-wave scattering length,
 the pion-nucleon $\sigma$ term $\sigma_{\pi N}$, 
and $m_0$ (from the value of the proton mass) at $\Opd$. 
  We do not   consider the isospin-odd $\pi N$ 
scattering length $a_{0+}^-$  since at this order 
is just given by $g_A$, in good agreement with experiment \cite{ulf}.
The $\sigma_{\pi N}$ term receives sizable higher order corrections 
from the  mesonic cloud  which are expected to be positive 
 and around $10$ MeV \cite{gasser2}.
 Since we evaluate  it just at $\Opd$, we enforce 
$\sigma_{\pi N}=$ 20, 30 or 40 MeV in the fits ($\sigma_{\pi N}=45\pm 8 $ MeV \cite{glsigma}). 
For the same reason, we enforce $m_0=$  0.7 or 0.8 GeV. 
We also include the value 
$a_{0+}^+=-(1\pm 1) \cdot 10^{-2}\,m_\pi^{-1}$ in the fit procedure. 
This value  results after considering its experimental one 
 $a_{0+}^+=-(0.25\pm 0.49) \cdot 10^{-2}\,m_\pi^{-1}$ \cite{schroder} 
and the theoretical expectation of
  positive $\Opt$ corrections around $+1\cdot 10^{-2}\,m_\pi^{-1}$ 
  from unitarity \cite{ulf}.   We
 stress that for all the fits we minimize strictly the $\chi^2$, 
that is, each data point is   weighted according to 
its experimental error. We do not include the data
 from \cite{cibo} in the $\sigma(K^- p\rightarrow \pi^-\Sigma^+)$ 
cross section since they are incompatible with all the  other  data.
 
{\bf 5.} We typically find two classes of fits, namely,
class A,  which give rise to   $1s$ kaonic hydrogen
$\Delta E$ and $\Gamma$  around the DEAR measurement,
 and class B fits, which are at variance with the DEAR measurement
but close to the results derived from Martin's scattering
 lengths \cite{martin}.  

\begin{figure}[Ht] 
\psfrag{Gamma}{{\tiny $ \Gamma$(eV)}}
\psfrag{pl}{{\tiny{$\begin{array}{c} p_L \hbox{(MeV)} \end{array}$}}}
\psfrag{sig1}{{\tiny $\sigma(K^-\!\!p\!\!\rightarrow\!\! K^-\!p)$(mb)}}
\psfrag{sig2}{{\tiny 
$\sigma(K^-\!\!p\!\!\rightarrow\!\!\bar{K}^0\!n)$(mb)}}
\psfrag{sig3}{{\tiny 
$\sigma(K^-\!\!p\!\!\rightarrow\!\!\pi^+\!\Sigma^-)$(mb)}}
\psfrag{sig4}{{\tiny 
$\sigma(K^-\!\!p\!\!\rightarrow\!\!\pi^- \!\Sigma^+)$(mb)}}
\psfrag{sig5}{{\tiny 
$\sigma(K^-\!\!p\!\!\rightarrow\!\!\pi^0\! \Sigma^0)$(mb)}}
\psfrag{sig6}{{\tiny 
$\sigma(K^-\!\!p\!\!\rightarrow\!\! \pi^0\! \Lambda)$(mb)}}
\psfrag{sig7}{{\tiny $\pi\Sigma$ Events/10 MeV }}
\psfrag{E}{{\tiny E(GeV)}}
\centerline{\epsfig{file=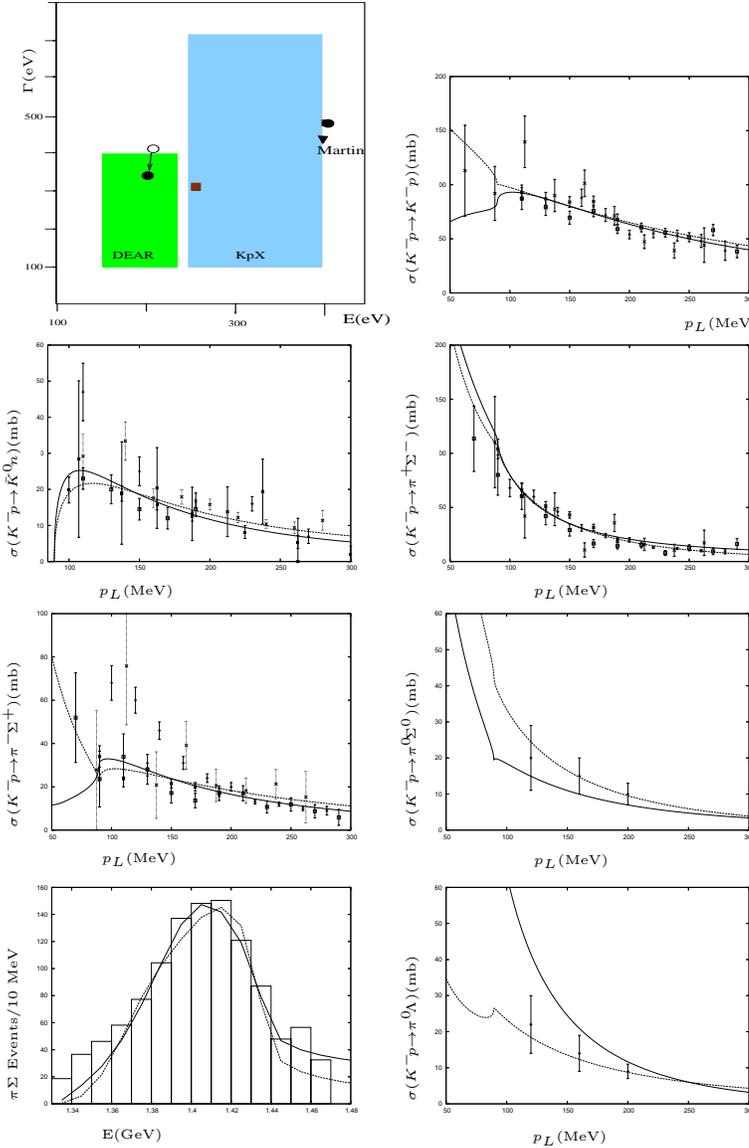,width=4.in,height=6.in}}
\caption[pilf]{\protect \small First panel: $1s$ kaonic hydrogen strong  energy shift and width.
In the rest, the solid lines correspond to the fit $A_4^+$  and  
the dashed ones to $B_4^+$. For further details see the text.
\label{fig:resul}}
\end{figure}

In Fig. \ref{fig:resul}, we show the shift and width of 
the $1s$ kaonic hydrogen state in the first panel and 
the cross sections and event distribution  data 
in the rest of panels.  The solid and dashed lines correspond to the fits 
with  $\sigma_{\pi N}=40$ MeV and 
$m_0=0.8$ GeV, called $A_4^+$ and $B_4^+$, respectively
--we discuss all the other fits in \cite{opv2}. 
Since the fit $B_4$ strongly disagrees  
with the DEAR measurement, we include in this fit
the KEK measurement and not the DEAR one.
In  the  first panel of Fig. \ref{fig:resul},
the solid circle on the left is for $A_4^+$ while the 
solid one on the right is for $B_4^+$. The empty 
 circle is obtained using the Deser
 formula \cite{deser} 
with the $K^-p$ scattering length from  $A_4^+$. 
We observe a gentle correction to the Deser formula result when using 
the expression including the isospin breaking corrections from \cite{akaki}. 
The downward triangle is the result of  using  Martin's scattering 
lengths \cite{martin} in \cite{akaki}.
The squares correspond to the fits    
with $\sigma_{\pi N}=30$ MeV and $m_0=0.8$ GeV, 
for details see \cite{opv2}. The isospin even $\pi N$ scattering 
length results always around $-1\cdot 10^{-2} m_\pi^{-1}$.
 The values for the ratios in (4) from  the fit $A_4^+(B_4^+)$
are $\gamma=2.36(2.36)$,  $R_c=0.628(0.655)$ and $R_n=0.172(0.195)$. 
 Both fits
reproduce data
remarkably well, even for higher energies  than included in the fit. 
The fitted parameters from $A_4^+(B_4^+)$ 
are, in GeV units: 
 $f=0.080(0.089)$, $b_0=-0.85(-0.32)$, $b_D=0.71(-0.10)$, 
$b_F=-0.04(-0.31)$, $b_1=0.60(-0.19)$,
 $b_2=1.07(-0.27)$, $b_3=-0.19(-0.15)$, $b_4=-1.25(-0.28)$, 
$\widetilde a_1=0.37(-0.05)$, $\widetilde a_2=1.14(-0.54)$,
 $\widetilde a_5=0.22(-1.08)$, $\widetilde a_7=0.00(-0.05)$, 
$\widetilde a_8=0.31(-0.54)$ and 
$\widetilde a_9=1.38(0.64)$. Notice 
that the $b_0$, $b_D$ and $b_F$ values from the fit $B_4^+$ are 
close to the values obtained from an
$\Opd$ CHPT analysis of baryon $1/2^+$ masses, while for 
the fit $A_4^+$ this is not the case. However,  we must stress that 
these couplings are not employed in the same formalism, UCHPT resums
 large contributions in this sector, 
 and then there is no reason 
why the values should be the same. Indeed, a pure  CHPT calculation of the lightest octet baryon 
masses is subject to huge higher order corrections and it is 
 always questionable \cite{gasser2}. The resulting 
$K^-p$ scattering length is $a_{K^-p}=(-0.51+i\,0.42)$ fm for 
the fit $A_4^+$  and $(-1.01+i\,0.80)$ fm for the fit $B_4^+$, i.e., 
a factor of two difference. 
Notice how the precise DEAR measurement
 places very severe constraints on the $\bar{K}N$ S-wave at threshold 
pointing  to a less  repulsive $K^-p$ interaction.
Indeed,  this is also reflected by the (two) $\Lambda(1405)$
pole positions which for the fit $A_4^+$ are at $(1321-i\,43.5)$ and 
$(1402-i\,39.6)$ MeV,  around 
30 to 40  MeV  lower than the fit $B_4^+$ ones
located at  $(1361-i\,29.9)$ and  $(1433-i\,31.7)$ MeV, respectively.
 This is  crucial  for $\bar{K}$-nucleus potential  calculations.
We therefore also confirm the presence of two rather narrow poles 
conforming the  $\Lambda(1405)$ \cite{team,osetprl}
with this higher order calculation. We agree with the  $K^-p$ scattering length in 
\cite{ivanov} although not for $a_0$ and $a_1$ separately. 
In the isospin limit, we get $a_0=(-1.23+i\,0.45)$ fm and 
$a_1=(0.98+i\,0.35)$ fm for the fit $A_4^+$ and 
$a_0=(-1.63+i\,0.81)$ fm and  $a_1=(-0.01+i\,0.54)$ fm for the fit 
$B_4^+$, where subindices refer to the $\bar{K}N$ isospin. 
 
 The  S-wave 
and P-wave phase shifts difference at the $\Xi^-$ mass has been recently 
determined from the 
measurement of the  $\Xi^-\rightarrow \Lambda \pi^-$   
decay parameters. The results are 
$\delta_P-\delta_S=(4.6\pm 1.4\pm 1.2)^{\hbox{\small o}}$ 
 \cite{hyperCP} and  
$(3.2\pm 5.3\pm 0.7 )^{\hbox{\small o}}$ \cite{e756}.
 Neglecting  the tiny P-wave phase shift \cite{ollerm2},
we obtain 
$2.5^{\hbox{\small o}}$ 
for the fit $A_4^+$ and $ 0.2^{\hbox{\small o}}$ for the fit $B_4^+$. 
Again the fit $A_4$ is in better agreement with the new 
$S=-1$ meson-baryon scattering data.   

{\bf 6.} In summary, 
we have presented a NLO analysis of S-wave $\bar{K}N$  scattering 
within UCHPT, that combines the second order SU(3) CHPT meson-baryon 
amplitudes with a dispersion
 relation for the inverse of a partial wave amplitude \cite{ollerm}.
 We have emphasized the  strong
 constraints that these precise data  imposes on 
the $\bar{K}N$ S-wave scattering amplitudes, implying 
 a less repulsive $K^-p$ interaction at threshold. 
This manifests in lower values for the two $\Lambda(1405)$ resonance 
poles  --whose presence we confirm at NLO.
As a novelty we find a class of fits (class A) which show consistency 
between  the DEAR and  scattering data, both old and  
new \cite{hyperCP,e756}.   
Further   exciting developments are foreseeable 
 with the DEAR/SIDDHARTA experiment \cite{sid} 
which aims at an eV level measurement of the shift 
and  width of kaonic hydrogen.

We would like to thank Daisuke Jido, Ulf-G. Mei{\ss}ner, Eulogio Oset and 
Akaki Rusetsky for useful discussions. 
J.A.O. is grateful to J\"urg Gasser for suggesting  him this problem. 
Financial support by the European Commission (EC)  
RTN Program Network ``EURIDICE'' 
 Contract No HPRN-CT-2002-00311, the HadronPhysics I3
Project (EC)  Contract No RII3-CT-2004-506078 (J.A.O.), by
MEC (Spain)  and FEDER (EC) Grants 
 No. FPA2004-03470 (J.A.O. and  M.V.) and
 FPA2003-09298-C02-01 (J.P.) are acknowledged.

\end{document}